\documentclass[aps,prc,twocolumn,floatfix,nofootinbib,showpacs,superscriptaddress]{revtex4-1}

\usepackage{graphicx}
\usepackage{dcolumn}
\usepackage{amsmath}
\usepackage{longtable}

\newcommand{\bea}{\begin{eqnarray}}
\newcommand{\eea}{\end{eqnarray}}

\usepackage{color}
\definecolor{purple}{rgb}{0.5,0,0.5}
\definecolor{blue}{rgb}{0.0,0,0.9}

\begin{document}
\title{
$\pi$- and $\rho$-mesons, and their diquark partners, from a contact interaction
}

\author{H.\,L.\,L.~Roberts}
\affiliation{Physics Division, Argonne National Laboratory, Argonne,
Illinois 60439, USA}
\affiliation{Institut f\"ur Kernphysik, Forschungszentrum J\"ulich, D-52425 J\"ulich, Germany}
\affiliation{Physics Department, University of California, Berkeley, California 94720, USA}

\author{A.~Bashir}
\affiliation{Instituto de F\'{\i}sica y Matem\'aticas,
Universidad Michoacana de San Nicol\'as de Hidalgo, Apartado Postal
2-82, Morelia, Michoac\'an 58040, Mexico}
\affiliation{Kavli Institute for Theoretical Physics China, CAS, Beijing 100190, China}

\author{L.\,X.~Guti\'errez-Guerrero}
\affiliation{Instituto de F\'{\i}sica y Matem\'aticas,
Universidad Michoacana de San Nicol\'as de Hidalgo, Apartado Postal
2-82, Morelia, Michoac\'an 58040, Mexico}

\author{C.\,D.~Roberts}
\affiliation{Physics Division, Argonne National Laboratory, Argonne, Illinois 60439, USA}
\affiliation{Institut f\"ur Kernphysik, Forschungszentrum J\"ulich, D-52425 J\"ulich, Germany}
\affiliation{Kavli Institute for Theoretical Physics China, CAS, Beijing 100190, China}
\affiliation{Department of Physics, Center for High Energy Physics and the State Key Laboratory of Nuclear Physics and Technology, Peking University, Beijing 100871, China}

\author{D.\,J.~Wilson}
\affiliation{Physics Division, Argonne National Laboratory, Argonne, Illinois 60439, USA}

\begin{abstract}
We present a unified Dyson-Schwinger equation treatment of static and electromagnetic properties of pseudoscalar and vector mesons, and scalar and axial-vector diquark correlations, based upon a vector-vector contact-interaction.  A basic motivation for this study is the need to document a comparison between the electromagnetic form factors of mesons and those diquarks which play a material role in nucleon structure.  This is an important step toward a unified description of meson and baryon form factors based on a single interaction.  A notable result, therefore, is the large degree of similarity between related meson and diquark form factors.  The simplicity of the interaction enables computation of the form factors at arbitrarily large spacelike $Q^2$, which enables us to expose a zero in the $\rho$-meson electric form factor at $z_Q^\rho \approx \surd 6 m_\rho$.  Notably, $r_\rho z_Q^\rho \approx r_{\rm D} z_Q^{\rm D}$, where $r_\rho$, $r_{\rm D}$ are, respectively, the electric radii of the $\rho$-meson and deuteron.
\end{abstract}
\maketitle

\section{Introduction}
In numerous respects, $\pi$- and $\rho$ mesons are the simplest bound-states to study in QCD.  That is, of course, supposing that the framework employed is Poincar\'e-covariant, capable of simultaneously describing light-quark confinement and dynamical chiral symmetry breaking (DCSB), and admits a symmetry-preserving truncation scheme.  All these features are required because, amongst many other things, the pion is the lightest hadron and QCD's Goldstone mode, the $\rho$-meson couples strongly to two pions and is an important part of the photon's vacuum polarisation, and modern facilities probe hadrons with momentum transfers far in excess of any reasonable constituent-quark-like mass-scale.

The Dyson-Schwinger equations (DSEs) \cite{Roberts:1994dr,Roberts:2000aa} provide an approach to hadron physics that is distinguished by its ability to satisfy these demands; and there is a large body of research that addresses $\pi$- and $\rho$-meson properties.  For example, the analysis of static properties is reported in Refs.\,\cite{Praschifka:1986nf,Roberts:1987xc,Cahill:1987qr,Praschifka:1989fd,%
Munczek:1991jb,Hollenberg:1992nj,Jain:1993qh,Mitchell:1994jj,Munczek:1994zz,Frank:1995uk,
Bender:1996bb,Burden:1996nh,Maris:1997hd,Maris:1997tm,Pichowsky:1999mu,Maris:1999nt,
Maris:2002yu,Holl:2004fr,Bhagwat:2007ha,Chang:2009zb,Fischer:2009jm} and of interactions in Refs.\,\cite{Roberts:1993ks,Roberts:1994hh,Alkofer:1995jx,Maris:1998hc,Kekez:1998rw,
Maris:1999bh,Bistrovic:1999dy,Maris:2000sk,Hecht:2000xa,Bicudo:2001jq,Maris:2002mz,
Jarecke:2002xd,Cotanch:2003xv,Maris:2004bp,Holl:2005vu,Bhagwat:2006pu,Holt:2010vj,
GutierrezGuerrero:2010md,Roberts:2010rn,Nguyen:2011jy}.  There is nevertheless a need to return to this theme; namely, a programme aimed at charting the interaction between light-quarks by explicating the impact of differing assumptions about the behaviour of the Bethe-Salpeter kernel on hadron elastic and transition form factors \cite{Aznauryan:2009da}.

To expose the connection we remark that in quantum field theory a baryon appears as a pole in a six-point quark Green function.  The pole's residue is proportional to the baryon's Faddeev amplitude, which is obtained from a Poincar\'e covariant Faddeev equation that sums all possible quantum field theoretical exchanges and interactions that can take place between three dressed-quarks.  A tractable truncation of the Faddeev equation is based \cite{Cahill:1988dx} on the observation that an interaction which describes mesons also generates diquark correlations in the colour-$\bar 3$ channel \cite{Cahill:1987qr}.  The dominant correlations for ground state octet and decuplet baryons are scalar ($0^+$) and axial-vector ($1^+$) diquarks because, for example, the associated mass-scales are smaller than the baryons' masses and their parity matches that of these baryons.  This is elucidated in Ref.\,\cite{Roberts:2011}.

At leading-order in a symmetry preserving truncation of the DSEs \cite{Munczek:1994zz,Bender:1996bb}, simple changes in the equations describing $\pi$- and $\rho$ mesons yield expressions that provide detailed information about the scalar and axial-vector diquarks; e.g., their masses \cite{Cahill:1987qr,Praschifka:1989fd,Burden:1996nh,Maris:2002yu,Roberts:2011,Roberts:2010hu}, and electromagnetic elastic \cite{Maris:2004bp} and transition form factors, which are critical elements in the computation of a baryon's kindred properties.  It is therefore natural to elucidate concurrently the properties of $\pi$- and $\rho$-mesons and those of the scalar and axial-vector diquark correlations because it opens the way to a unified, symmetry-preserving explanation of meson and baryon properties as they are predicted by a single interaction.  The potential of this approach is apparent in Refs.\,\cite{Eichmann:2008ae,Eichmann:2008ef} but it has yet to be fully realised.  For the present the best connection is provided by the less rigorous approach of Ref.\,\cite{Cloet:2008re}, which uses more parameters to express features of QCD but also predicts and describes simultaneously a larger array of phenomena \cite{deJager:2009xs,Puckett:2010ac,Riordan:2010id}.

Herein, as part of the programme outlined above, we describe results for a range of static and dynamic properties of these simplest $u/d$-mesons and -diquark-correlations as produced by a vector-vector current-current interaction that is mediated by a momentum-independent boson propagator; i.e., by the symmetry-preserving regularisation of a contact interaction.  Given the large body of work based on QCD-like vector-boson propagation that is already available, this study will provide numerous points for comparison and contrast that are relevant to existing and planned experiments.

In Sec.\,\ref{sec:model} we describe a symmetry-preserving regularisation and DSE-formulation of the contact interaction, following Refs.\,\cite{GutierrezGuerrero:2010md,Roberts:2010rn,Roberts:2011}.  Our scheme is such that confinement is manifest, and chiral symmetry and the pattern by which it is broken are veraciously represented.  In addition to the current-quark mass, the model has two parameters.
In Sec.\,\ref{sec:formfactors} we describe results for $\pi$- and $\rho$-meson electromagnetic elastic and transition form factors, computed using the rainbow-ladder truncation of the DSEs; with the analogous discussion of diquark correlations reported in Sec.\ref{sec:formfactorsdiquarks}.
Section~\ref{sec:epilogue} provides a summary and perspective.

\section{Contact vector-current-current interaction}
\label{sec:model}
\subsection{Gap equation}
The typical starting point for a DSE study of hadron phenomena is the dressed-quark propagator, which is obtained from the gap equation:
\begin{eqnarray}
\nonumber \lefteqn{S(p)^{-1}= i\gamma\cdot p + m}\\
&&+ \int \! \frac{d^4q}{(2\pi)^4} g^2 D_{\mu\nu}(p-q) \frac{\lambda^a}{2}\gamma_\mu S(q) \frac{\lambda^a}{2}\Gamma_\nu(q,p) ,
\label{gendse}
\end{eqnarray}
wherein $m$ is the Lagrangian current-quark mass, $D_{\mu\nu}$ is the vector-boson propagator and $\Gamma_\nu$ is the quark--vector-boson vertex.  Much is now known about $D_{\mu\nu}$ in QCD \cite{Kamleh:2007ud,Dudal:2008rm,Cucchieri:2009zt,RodriguezQuintero:2010wy} and nonperturbative information is accumulating on $\Gamma_\nu$ \cite{Chang:2009zb,Skullerud:2003qu,Kizilersu:2009kg,Chang:2010jq,Chang:2010hb}.

However, our goal is to build a stock of material that can be used to identify unambiguous signals in experiment for the pointwise behaviour of: the interaction between light-quarks; the light-quarks' mass-function; and other similar quantities.  Whilst these are particular qualities, taken together they will enable a characterisation of the nonperturbative behaviour of the theory underlying strong interaction phenomena \cite{Holt:2010vj,Aznauryan:2009da}.  We therefore elucidate predictions following from the assumption\footnote{This choice is the antithetical complement to that proposed in Ref.\,\protect\cite{Munczek:1983dx}; i.e., a $\delta$-function in four-momentum space, which is confining because it provides a strong interaction that is independent of separation, $x^2$.}
\begin{equation}
\label{njlgluon}
g^2 D_{\mu \nu}(p-q) = \delta_{\mu \nu} \frac{1}{m_G^2}\,,
\end{equation}
where $m_G$ is a gluon mass-scale, and proceed by embedding this interaction in a rainbow-ladder truncation of the DSEs, which is the leading-order in the most widely used, symmetry-preserving truncation scheme \cite{Bender:1996bb}.  This means
\begin{equation}
\Gamma_{\nu}(p,q) =\gamma_{\nu}
\end{equation}
in Eq.\,(\ref{gendse}) and in the subsequent construction of the Bethe-Salpeter kernels.

With this kernel the gap equation becomes
\begin{equation}
 S^{-1}(p) =  i \gamma \cdot p + m +  \frac{4}{3}\frac{1}{m_G^2} \int\!\frac{d^4 q}{(2\pi)^4} \,
\gamma_{\mu} \, S(q) \, \gamma_{\mu}\,,   \label{gap-1}
\end{equation}
an equation in which the integral possesses a quadratic divergence, even in the chiral limit.  If the divergence is regularised in a Poincar\'e covariant manner, then the solution is
\begin{equation}
\label{genS}
S(p)^{-1} = i \gamma\cdot p + M\,,
\end{equation}
where $M$ is momentum-independent and determined by
\begin{equation}
M = m + \frac{M}{3\pi^2 m_G^2} \int_0^\infty \!ds \, s\, \frac{1}{s+M^2}\,.
\end{equation}

One must specify a regularisation procedure in order to proceed.  We write \cite{Ebert:1996vx}
\begin{eqnarray}
\nonumber
\frac{1}{s+M^2} & = & \int_0^\infty d\tau\,{\rm e}^{-\tau (s+M^2)} \\
& \rightarrow & \int_{\tau_{\rm uv}^2}^{\tau_{\rm ir}^2} d\tau\,{\rm e}^{-\tau (s+M^2)}
\label{RegC}\\
& & =
\frac{{\rm e}^{- (s+M^2)\tau_{\rm uv}^2}-e^{-(s+M^2) \tau_{\rm ir}^2}}{s+M^2} \,, \label{ExplicitRS}
\end{eqnarray}
where $\tau_{\rm ir,uv}$ are, respectively, infrared and ultraviolet regulators.  It is apparent from Eq.\,(\ref{ExplicitRS}) that a nonzero value of $\tau_{\rm ir}=:1/\Lambda_{\rm ir}$ implements confinement by ensuring the absence of quark production thresholds \cite{Krein:1990sf,Roberts:2007ji}.  We note that since Eq.\,(\ref{njlgluon}) does not define a renormalisable theory,  $\Lambda_{\rm uv}:=1/\tau_{\rm uv}$ cannot be removed but instead plays a dynamical role and sets the scale of all dimensioned quantities.  The gap equation can now be written
\begin{equation}
M = m +  \frac{M}{3\pi^2 m_G^2} \,{\cal C}^{\rm iu}(M^2)\,,
\label{gapactual}
\end{equation}
where ${\cal C}^{\rm iu}(M^2)/M^2 = \Gamma(-1,M^2 \tau_{\rm uv}^2) - \Gamma(-1,M^2 \tau_{\rm ir}^2)$, with $\Gamma(\alpha,y)$ being the incomplete gamma-function.

\subsection{Point-meson Bethe-Salpeter equation}
In rainbow-ladder truncation, with the interaction in Eq.\,(\ref{njlgluon}), the homogeneous Bethe-Salpeter equation for a colour-singlet meson is
\begin{equation}
\Gamma(k;P) = -\frac{4}{3}\frac{1}{m_G^2} \int \! \frac{d^4q}{(2\pi)^4}\, \gamma_\mu \chi(q;P)\gamma_\mu \,,
\label{genbse}
\end{equation}
where $\chi(q;P) = S(q+P)\Gamma(q;P)S(q)$ and $\Gamma(q;P)$ is the meson's Bethe-Salpeter amplitude.  Since the integrand does not depend on the external relative-momentum, $k$, then a symmetry-preserving regularisation of Eq.\,(\ref{genbse}) will yield solutions that are independent of $k$.  It follows that if the interaction in Eq.\,(\ref{njlgluon}) produces bound states, then the relative momentum between the bound-state's constituents can assume any value with equal probability.  This is the defining characteristic of a pointlike composite particle.

With a dependence on the relative momentum forbidden by the interaction, the pseudoscalar and vector Bethe-Salpeter amplitudes take the general form\footnote{We assume isospin symmetry throughout and hence do not include the Pauli isospin matrices explicitly.} \cite{LlewellynSmith:1969az}
\begin{eqnarray}
\Gamma^\pi(P) &= &  i \gamma_5 E_\pi(P) + \frac{1}{M} \gamma_5 \gamma\cdot P F_\pi(P) \,,\\
\Gamma_\mu^\rho(P) & = & \gamma^T_\mu E_\rho(P)+ \frac{1}{M} \sigma_{\mu\nu} P_\nu F_\rho(P)\,, \label{rhobsa}
\end{eqnarray}
where $P_\mu \gamma^T_\mu = 0$ and $\gamma^T_\mu+\gamma^L_\mu=\gamma_\mu$.  We observe that
\begin{equation}
\label{Frhozero}
F_\rho(P) \stackrel{\mbox{\footnotesize ladder}}{\equiv} 0\,.
\end{equation}
However, it should be borne in mind that this is an artefact of the rainbow-ladder truncation; viz., even using Eq.\,(\ref{njlgluon}), $F_\rho(P)\neq 0$ in any symmetry-preserving truncation that goes beyond this leading-order \cite{Bender:1996bb}.  We will see that the accident expressed in Eq.\,(\ref{Frhozero}) has material consequences.

\subsection{Ward-Takahashi identities}
\label{sec:WTI}
No study of $\pi$- or $\rho$-meson observables is meaningful unless it ensures expressly that the vector and axial-vector Ward-Takahashi identities are satisfied.  The $m=0$ axial-vector identity states 
\begin{equation}
\label{avwti}
P_\mu \Gamma_{5\mu}(k_+,k) = S^{-1}(k_+) i \gamma_5 + i \gamma_5 S^{-1}(k)\,,
\end{equation}
where $\Gamma_{5\mu}(k_+,k)$ is the axial-vector vertex, which is determined by
\begin{equation}
\Gamma_{5\mu}(k_+,k) =\gamma_5 \gamma_\mu
- \frac{4}{3}\frac{1}{m_G^2} \int\frac{d^4q}{(2\pi)^4} \, \gamma_\alpha \chi_{5\mu}(q_+,q) \gamma_\alpha\,. \label{aveqn}
\end{equation}

We must therefore implement a regularisation that maintains Eq.\,(\ref{avwti}).  This requirement is readily found to entail the following two chiral limit identities \cite{GutierrezGuerrero:2010md}:
\begin{eqnarray}
\label{Mavwti}
M & = & \frac{8}{3}\frac{M}{m_g^2} \int\! \frac{d^4q}{(2\pi)^4} \left[ \frac{1}{q^2+M^2} +  \frac{1}{q_+^2+M^2}\right],\\
\label{0avwti}
0 & = & \int\! \frac{d^4q}{(2\pi)^4} \left[ \frac{P\cdot q_+}{q_+^2+M^2} -  \frac{P\cdot q}{q^2+M^2}\right]\,,
\end{eqnarray}
which must be satisfied after regularisation.  Analysing the integrands using a Feynman parametrisation, one arrives at the follow identities for $P^2=0=m$:
\begin{eqnarray}
 M & = &  \frac{16}{3}\frac{M}{m_G^2} \int\! \frac{d^4q}{(2\pi)^4} \frac{1}{[q^2+M^2]}, \label{avwtiMc}\\
0 & = & \int\! \frac{d^4q}{(2\pi)^4} \frac{\frac{1}{2} q^2 + M^2 }{[q^2+M^2]^2} \label{avwtiAc}.
\end{eqnarray}

Equation\,(\ref{avwtiMc}) is just the chiral-limit gap equation.  Hence it requires nothing new of the regularisation scheme.   On the other hand, Eq.\,(\ref{avwtiAc}) states that the axial-vector Ward-Takahashi identity is satisfied if, and only if, the model is regularised so as to ensure there are no quadratic or logarithmic divergences.  Unsurprisingly, these are the just the circumstances under which a shift in integration variables is permitted, an operation required in order to prove Eq.\,(\ref{avwti}).

It is notable, too, that Eq.\,(\ref{avwti}) is valid for arbitrary $P$.  In fact its corollary, Eq.\,(\ref{Mavwti}), may be used to demonstrate that in the chiral limit the two-flavour scalar-meson rainbow-ladder truncation of the contact-interaction DSEs produces a bound-state with mass $m_\sigma = 2 \,M$ \cite{Roberts:2011,Roberts:2010gh}.  In the presence of a momentum-dependent dressed-quark mass function, one could reverse this association and define a chiral-limit dressed-quark constituent-mass as one-half the mass of the lightest rainbow-ladder scalar meson.  This procedure yields $M^0 \simeq 0.3\,$GeV, as may readily be determined from Ref.\,\cite{Roberts:2000aa}.

The second corollary, Eq.\,(\ref{0avwti}), entails
\begin{equation}
0 = \int_0^1d\alpha \,
\left[ {\cal C}^{\rm iu}(\omega(M^2,\alpha,P^2))  + \, {\cal C}^{\rm iu}_1(\omega(M^2,\alpha,P^2))\right], \label{avwtiP}
\end{equation}
with
\begin{eqnarray}
\label{eq:omega}
\omega(M^2,\alpha,P^2) &=& M^2 + \alpha(1-\alpha) P^2\,,\\
\nonumber
{\cal C}^{\rm iu}_1(z) &=& - z (d/dz){\cal C}^{iu}(z) \\
&= & z\left[ \Gamma(0,M^2 r_{\rm uv}^2)-\Gamma(0,M^2 r_{\rm ir}^2)\right] .
\label{eq:C1}
\end{eqnarray}

The vector Ward-Takahashi identity
\begin{equation}
\label{VWTI}
P_\mu i\Gamma^\gamma_\mu(k_+,k) = S^{-1}(k_+) - S^{-1}(k)\,,
\end{equation}
wherein $\Gamma^\gamma_\mu$ is the dressed-quark-photon vertex, is crucial for a sensible study of electromagnetic form factors \cite{Roberts:1994hh}.  Ideally, the vertex needs to be dressed at a level consistent with the truncation used to compute the bound-state's Bethe-Salpeter amplitude \cite{Maris:1999bh}.  In our case this means the vertex should be determined from the following inhomogeneous Bethe-Salpeter equation:
\begin{equation}
\label{GammaQeq}
\Gamma_\mu(Q) = \gamma_\mu - \frac{4}{3} \frac{1}{m_G^2} \int \frac{d^4 q}{(2\pi)^4} \, \gamma_\alpha \chi_\mu(q_+,q) \gamma_\alpha\,,
\end{equation}
where $\chi_\mu(q_+,q) = S(q+P) \Gamma_\mu (Q) S(q)$.  Owing to the momentum-independent nature of the interaction kernel, the general form of the solution is
\begin{equation}
\label{GammaQ}
\Gamma_\mu(Q) = \gamma^T_\mu P_T(Q^2) + \gamma_\mu^L P_L(Q^2)\,,
\end{equation}
where $Q_\mu \gamma^T_\mu = 0$ and $\gamma^T_\mu+\gamma^L_\mu=\gamma_\mu$.  This simplicity doesn't survive with a more sophisticated interaction nor with Eq.\,(\ref{njlgluon}) beyond rainbow-ladder truncation \cite{Roberts:2011}.

Inserting Eq.\,(\ref{GammaQ}) into Eq.\,(\ref{GammaQeq}), one readily obtains
\begin{equation}
\label{PL0}
P_L(Q^2)= 1\,,
\end{equation}
owing to Eq.\,(\ref{0avwti}).  Using this same identity, one finds
\begin{equation}
\label{PTQ2}
P_T(Q^2)= \frac{1}{1+K_\gamma(Q^2)},
\end{equation}
with ($\overline{\cal C}_1(z) = {\cal C}_1(z)/z$)
\begin{eqnarray}
\nonumber
\lefteqn{K_\gamma(Q^2) = \frac{1}{3\pi^2m_G^2}}\\
& & \times \int_0^1d\alpha\, \alpha(1-\alpha) Q^2\,  \overline{\cal C}^{iu}_1(\omega(M^2,\alpha,Q^2))\,. \label{Kgamma}
\end{eqnarray}
Plainly,
\begin{equation}
\label{PT0}
P_T(Q^2=0)=1\,,
\end{equation}
so that at $Q^2=0$ in the rainbow-ladder treatment of the interaction in Eq.\,(\ref{njlgluon}) the dressed-quark-photon vertex is equal to the bare vertex.\footnote{Equations~(\ref{PL0}), (\ref{PT0}) guarantee a massless photon and show that our regularisation also ensures preservation of the Ward-Takahashi identity for the photon vacuum polarisation \protect\cite{Burden:1991uh}.}

However, this is not true for $Q^2\neq 0$.  In fact the transverse part of the dressed-quark-photon vertex will display a pole at that $Q^2<0$ for which
\begin{equation}
\label{rhobse}
1+K_\gamma(Q^2)=0\,.
\end{equation}
This is just the model's Bethe-Salpeter equation for the ground-state vector meson.

\begin{figure}[t] 
\centerline{\includegraphics[clip,width=0.45\textwidth]{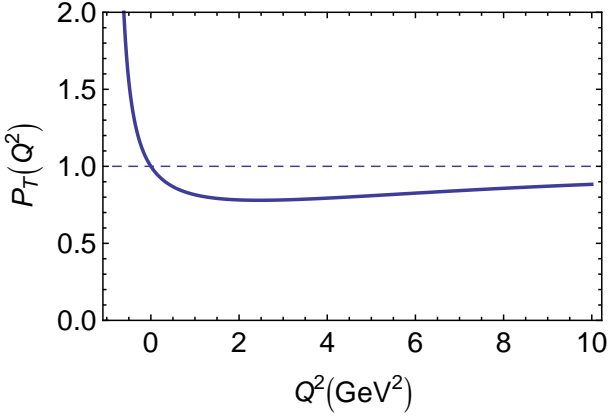}}
\caption{\label{qgammavertex} (Color online) Dressing function for the transverse piece of the quark-photon vertex; viz., $P_T(Q^2)$ in Eq.\,(\protect\ref{PTQ2}).
}
\end{figure}

In Fig.\,\ref{qgammavertex} we depict the function that dresses the transverse part of the quark-photon vertex.  The pole associated with the ground-state vector meson is clear.  Another important feature is the behaviour at large spacelike-$Q^2$; namely,  $P_T(Q^2) \to 1^-$ as $Q^2\to \infty$.  This is the statement that a dressed-quark is pointlike to a large-$Q^2$ probe.  The same is true in QCD, up to the logarithmic corrections which are characteristic of an asymptotically free theory \cite{Maris:1999bh}.

\subsection{Bethe-Salpeter kernels for $\pi$ and $\rho$}
At this point we can write the explicit form of Eq.\,(\ref{genbse}) for the pion:
\begin{equation}
\label{bsefinal0}
\left[
\begin{array}{c}
E_\pi(P)\\
F_\pi(P)
\end{array}
\right]
= \frac{1}{3\pi^2 m_G^2}
\left[
\begin{array}{cc}
{\cal K}_{EE} & {\cal K}_{EF} \\
{\cal K}_{FE} & {\cal K}_{FF}
\end{array}\right]
\left[\begin{array}{c}
E_\pi(P)\\
F_\pi(P)
\end{array}
\right],
\end{equation}
where
\begin{eqnarray}
\nonumber
{\cal K}_{EE} &= &\int_0^1d\alpha \left[ {\cal C}^{\rm iu}(\omega(M^2,\alpha,-m_\pi^2))\right.\\
&& \left.  + 2 \alpha(1-\alpha) \, m_\pi^2 \, \overline{\cal C}^{\rm iu}_1(\omega(M^2,\alpha,-m_\pi^2))\right],\\
{\cal K}_{EF} &=& -m_\pi^2 \int_0^1d\alpha\, \overline{\cal C}^{\rm iu}_1(\omega(M^2,\alpha,-m_\pi^2)), \\
{\cal K}_{FE} &=& \frac{1}{2} M^2 \int_0^1d\alpha\, \overline{\cal C}^{\rm iu}_1(\omega(M^2,\alpha,-m_\pi^2)),\\
{\cal K}_{FF} &=& - 2 {\cal K}_{FE}\,. \label{KpiFF}
\end{eqnarray}
This is an eigenvalue problem for the pion mass-squared, $m_\pi^2$.  NB.\ We used Eq.\,(\ref{avwtiP}) to arrive at Eq.\,(\ref{KpiFF}).

The explicit form of Eq.\,(\ref{genbse}) for the $\rho$-meson, whose solution yields its mass-squared, is
\begin{equation}
1+ K_\gamma(-m_\rho^2) = 0\,,
\end{equation}
where $K_\gamma$ is given in Eq.\,(\ref{Kgamma}).

In the computation of observables, one must use the canonically-normalised Bethe-Salpeter amplitudes.  For the rainbow-ladder pion this means that $\Gamma_\pi$ is rescaled to satisfy
\begin{equation}
P_\mu = N_c\, {\rm tr} \int\! \frac{d^4q}{(2\pi)^4}\Gamma_\pi(-P)
 \frac{\partial}{\partial P_\mu} S(q+P) \, \Gamma_\pi(P)\, S(q)\,, \label{Ndef}
\end{equation}
which, in the chiral limit, becomes
\begin{equation}
1 = \frac{N_c}{4\pi^2} \frac{1}{M^2} \, {\cal C}_1(M^2;\tau_{\rm ir}^2,\tau_{\rm uv}^2)
E_\pi [ E_\pi - 2 F_\pi].
\label{Norm0}
\end{equation}
For the rainbow-ladder $\rho$-meson, on the other hand, the vector meson analogue of Eq.\,(\ref{Ndef}) requires that
\begin{equation}
\label{Erhonorm}
\frac{1}{E_\rho^2} = - 9 m_G^2 \left. \frac{d}{dz} K_\gamma(z)\right|_{z=-m_\rho^2}.
\end{equation}

In terms of the canonically normalised Bethe-Salpeter amplitudes, the leptonic decay constants of the $\pi$- and $\rho$-mesons are respectively given by
\begin{eqnarray}
\label{fpim}
f_\pi & = & \frac{1}{M}\frac{3}{2\pi^2} \,[ E_\pi - 2 F_\pi] \,{\cal K}_{FE}^{P^2=-m_\pi^2} ,\\
f_\rho & = & - \frac{9}{2} \, \frac{E_\rho}{m_\rho} \, K_\gamma(-m_\rho^2)\,.
\end{eqnarray}

Another important low-energy property is the in-pion condensate\footnote{There is an analogous in-$\rho$-meson condensate but that will be discussed elsewhere.}
\begin{equation}
\label{kpim}
\kappa_\pi =  f_\pi \frac{3}{4\pi^2} [ E_\pi {\cal K}_{EE}^{P^2=-m_\pi^2} + F_\pi \,{\cal K}_{EF}^{P^2=-m_\pi^2} ]\,.
\end{equation}
In the chiral limit $\kappa_\pi \to \kappa_\pi^0 = -\langle \bar q q \rangle$; i.e., the so-called vacuum quark condensate \cite{Brodsky:2010xf}.  Moreover, in this limit, too, one can readily verify that \cite{GutierrezGuerrero:2010md}
\begin{equation}
\label{GTE}
E_\pi \stackrel{m=0}{=} \frac{M}{f_\pi}\,,
\end{equation}
which is a particular case of one of the Goldberger-Treiman relations proved in Ref.\,\cite{Maris:1997hd}, and  $F_\pi(P=0)$ satisfies a similar identity.

\section{\mbox{\boldmath $\pi$} and \mbox{\boldmath $\rho$} elastic and transition form factors}
\label{sec:formfactors}
In order to compute the form factors we need to fix the model's two parameters; namely, $m_G$ and $\Lambda_{\rm uv}$.\footnote{We fix $\Lambda_{\rm ir}=0.24\,$GeV\,$
\approx \Lambda_{\rm QCD}$ since $r_{\rm QCD}:=1/\Lambda_{\rm QCD}\approx 0.8\,$fm is a length-scale typical of confinement.}  We do that by performing a least-squares fit in the chiral limit to $M^0=0.40\,$GeV, $\kappa_\pi^0= (0.22\,{\rm GeV})^3$, $f_\pi^0=0.088\,$GeV, $m_\rho^0= 0.78\,$GeV and $f_\rho^0 = 0.15\,$GeV.  This procedure yields the results in Table~\ref{Table:static}.

\begin{table}[t]
\caption{Results obtained with (in GeV) $m_G=0.132\,$, $\Lambda_{\rm ir} = 0.24\,$, $\Lambda_{\rm uv}=0.905$, which yield a root-mean-square relative-error of 13\% in comparison with our specified goals for the observables. Dimensioned quantities are listed in GeV.
\label{Table:static}
}
\begin{center}
\begin{tabular*}
{\hsize}
{
l@{\extracolsep{0ptplus1fil}}
|c@{\extracolsep{0ptplus1fil}}
c@{\extracolsep{0ptplus1fil}}
c@{\extracolsep{0ptplus1fil}}
|c@{\extracolsep{0ptplus1fil}}
c@{\extracolsep{0ptplus1fil}}
c@{\extracolsep{0ptplus1fil}}
c@{\extracolsep{0ptplus1fil}}
c@{\extracolsep{0ptplus1fil}}
c@{\extracolsep{0ptplus1fil}}}\hline
$m$ & $E_\pi$ & $F_\pi$ & $E_\rho$ & $M$ & $\kappa_\pi^{1/3}$ & $m_\pi$ & $m_\rho$ & $f_\pi$ & $f_\rho$ \\\hline
0 & 3.568 & 0.459 & 1.520 & 0.358 & 0.241 & 0\,~~~~~ & 0.919 & 0.100 & 0.130\rule{0ex}{2.5ex}\\
0.007 & 3.639 & 0.481 & 1.531 & 0.368 & 0.243 & 0.140 & 0.928 & 0.101 & 0.129\\\hline
\end{tabular*}
\end{center}
\end{table}

\subsection{\mbox{\boldmath $\pi$}-meson elastic form factors}
\begin{figure}[t] 
\centerline{\includegraphics[clip,width=0.45\textwidth]{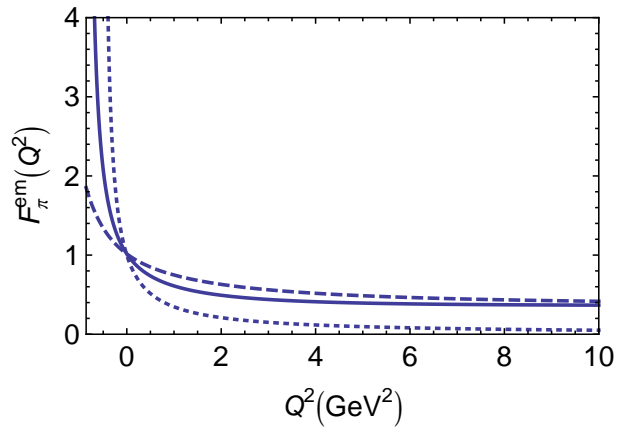}}
\caption{\label{figrhoFpi} (Color online) $F^{\rm em}_{\pi}(Q^2)$ computed in rainbow-ladder truncation from the interaction in Eq.\,(\protect\ref{njlgluon}): \emph{solid curve} -- fully consistent, i.e., with a dressed-quark-photon vertex so that the $\rho$-pole appears; and \emph{dashed curve} -- computed using a bare quark-photon vertex.  \emph{Dotted curve} -- fit to the result in Ref.\,\protect\cite{Maris:2000sk}, which also included a consistently-dressed quark-photon vertex and serves to illustrate the trend of contemporary data.
}
\end{figure}

We are solving the interaction of Eq.\,(\ref{njlgluon}) in the rainbow-ladder truncation; i.e., at leading-order in the nonperturbative symmetry-preserving truncation of Refs.\,\cite{Munczek:1994zz,Bender:1996bb}.  At this order the generalised impulse approximation is computed for three-point scattering processes \cite{Roberts:1994hh}, such as elastic form factors.  An analysis of the associated triangle diagram yields the formulae in Sec.\,\ref{app:FpiQ2} and the computed result is depicted in Fig.\,\ref{figrhoFpi}.  Two features are immediately apparent; viz., the pole associated with the $\rho$-meson at timelike momentum, which is a consequence of dressing the quark-photon vertex; and a momentum-independent interaction produces $F_\pi(Q^2) = \,$constant as $Q^2\to \infty$.  The following function is a valid interpolation of the full result on the domain shown:
\begin{equation}
F_\pi^{\rm em}(Q^2) \stackrel{\rm interpolation}{=}
\frac{1+0.33 \, Q^2 +0.024\, Q^4}{1+1.20 \, Q^2 + 0.053 \, Q^4}
\end{equation}

In Table\,\ref{Table:radii} we report the pion charge radius:
\begin{equation}
\label{rpidefn}
r_\pi^2 = - 6 \left. \frac{d}{dQ^2} F_\pi(Q^2) \right|_{Q^2=0}.
\end{equation}
The result is less than experiment ($r_\pi=0.672\pm 0.008\,$fm \cite{Nakamura:2010zzi}).  This owes in small part to our omission of pseudoscalar meson rescattering effects \cite{Alkofer:1993gu} but more to the large value we obtain for the $\rho$-meson's mass.  It cannot be remedied in our symmetry-preserving rainbow-ladder treatment of Eq.\,(\ref{njlgluon}) because all dimensioned quantities are too closely tied to the value of $M$.  An interaction which preserves the one-loop renormalisation group behaviour of QCD \cite{Maris:1997tm,Maris:1999nt} provides decoupling between the values of ultraviolet and infrared phenomena, such as $m_\rho$ and $\kappa_\pi$.

\begin{table}[bt]
\caption{\emph{Row 1}: Form factor radii (in fm), and magnetic and quadrupole moments for the $\rho$-meson, $G_M^\rho(Q^2=0)$ and $G_Q^\rho(Q^2=0)$ respectively, computed with (in GeV) $m=0.007$, $m_G=0.132\,$, $\Lambda_{\rm ir} = 0.24\,$, $\Lambda_{\rm uv}=0.905$.
For a structureless vector meson, $\mu=2$ and ${\cal Q}=-1$ \protect\cite{Brodsky:1992px}.
The next four rows list results reported elsewhere.
Experimentally, $r_\pi=0.672\pm 0.008\,$fm \cite{Nakamura:2010zzi}.
(NB.\ None of the quoted computations included contributions from nonresonant pseudoscalar-meson final-state interactions and hence agreement with the experimental value of $r_\pi$ should be seen as a defect of the associated model \protect\cite{Alkofer:1993gu}.  The nature of this flaw is understood within the DSE context, where such contributions can be viewed as computable corrections to the rainbow-ladder truncation \protect\cite{Eichmann:2008ae}.)
The last two lines report results for the scalar and axial-vector diquark correlations. Here the magnetic and quadrupole moments should be multiplied by the relevant charge factor; viz., $e_{\{uu\}}=\frac{4}{3}$, $e_{\{ud\}}=\frac{1}{3}$ and $e_{\{dd\}}=-\frac{2}{3}$.
\label{Table:radii}
}
\begin{center}
\begin{tabular*}
{\hsize}
{
l@{\extracolsep{0ptplus1fil}}
c@{\extracolsep{0ptplus1fil}}
c@{\extracolsep{0ptplus1fil}}
c@{\extracolsep{0ptplus1fil}}
c@{\extracolsep{0ptplus1fil}}
c@{\extracolsep{0ptplus1fil}}
c@{\extracolsep{0ptplus1fil}}}\hline
& $r_\pi$ & $r_\rho^E $ & $r_\rho^M$ & $r_\rho^E$ & $\mu_\rho$ & ${\cal Q}_\rho$ \\
This work & 0.45 & 0.56 & 0.51 & 0.51 & 2.11 & -0.85 \\
Ref.\,\protect\cite{Bhagwat:2006pu} & 0.66 & 0.73 &      &      & 2.01 & -0.41 \\
Refs.\,\protect\cite{Burden:1995ve,Hawes:1998bz}
    & 0.56 & 0.61 &  &  & 2.69 & -0.84 \\
Refs.\,\protect\cite{deMelo:1997cb,deMelo:1997hh}
    & 0.66 & 0.61 &  &  & 2.14 & -0.79 \\
Refs.\,\protect\cite{Choi:1997iq,Choi:2004ww}
    & 0.66 & 0.52 &  &  & 1.92 & -0.43 \\\hline
& $r_{0^+}$ & $r_{1^+}^E $ & $r_{1^+}^M$ & $r_{1^+}^E$ & $\mu_{1^+}$ & ${\cal Q}_{1^+}$ \\
This work    & 0.49 &  0.55 &  0.51 & 0.51 & 2.13 & -0.81 \\
Ref.\.\protect\cite{Maris:2004bp} & 0.71 &    &   &   &   &   \\\hline
\end{tabular*}
\end{center}
\end{table}

\subsection{\mbox{\boldmath $\rho$}-meson elastic form factors}
The $J^{PC}=1^{--}$ $\rho$-meson has three elastic form factors and we follow Ref.\,\cite{Bhagwat:2006pu} in defining them.  Denoting the incoming photon momentum by $Q$, and the incoming and outgoing $\rho$-meson momenta by $p^i= K-Q/2$ and $p^f= K+Q/2$, then $K\cdot Q=0$, $K^2+Q^2/4= -m_\rho^2$ and the $\rho$-$\gamma$-$\rho$ vertex can be expressed:
\begin{eqnarray}
\label{Lambdarho}
\Lambda_{\lambda,\mu\nu}(K,Q) & = & \sum_{j=1}^3 T_{\lambda,\mu\nu}^j(K,Q) \, F_j(Q^2)\,,\\
T_{\lambda,\mu\nu}^1(K,Q) & = & 2 K_\lambda\, {\cal P}^T_{\mu\alpha}(p^i) \, {\cal P}^T_{\alpha\nu}(p^f)\,,\\
\nonumber
T_{\lambda,\mu\nu}^2(K,Q) & = & \left[Q_\mu - p^i_\mu \frac{Q^2}{2 m_\rho^2}\right] {\cal P}^T_{\lambda\nu}(p^f) \\
&& - \left[Q_\nu + p^f_\nu \frac{Q^2}{2 m_\rho^2}\right] {\cal P}^T_{\lambda\mu}(p^i)\,, \\
\nonumber
T_{\lambda,\mu\nu}^3(K,Q) & = & \frac{K_\lambda}{m_\rho^2}\, \left[Q_\mu - p^i_\mu \frac{Q^2}{2 m_\rho^2}\right] \left[Q_\nu + p^f_\nu \frac{Q^2}{2 m_\rho^2}\right] \,,\\
&&
\end{eqnarray}
where ${\cal P}^T_{\mu\nu}(p) = \delta_{\mu\nu} - p_\mu p_\nu/p^2$.  A symmetry-preserving regularisation scheme is essential here so that the following Ward-Takahashi identities are preserved throughout the analysis:
\begin{eqnarray}
Q_\lambda \Lambda_{\lambda,\mu\nu}(K,Q) &=& 0\\
p^i_\mu \Lambda_{\lambda,\mu\nu}(K,Q) &=& 0 = p^f_\nu \Lambda_{\lambda,\mu\nu}(K,Q)\,.
\end{eqnarray}

The electric, magnetic and quadrupole form factors are constructed as follows:
\begin{eqnarray}
G_E(Q^2) & = & F_1(Q^2)+\frac{2}{3} \eta G_Q(Q^2)\,,\\
G_M(Q^2) & = & - F_2(Q^2)\,,\\
%
G_Q(Q^2) & = & F_1(Q^2) + F_2(Q^2) + \left[1+\eta\right] F_3(Q^2)\,,
\end{eqnarray}
where $\eta=Q^2/[4 m_\rho^2]$.  In the limit $Q^2\to 0$, these form factors define the charge, and magnetic and quadrupole moments of the $\rho$-meson; viz.,
\begin{eqnarray}
\label{chargenorm}
G_E^\rho(Q^2=0) & = & 1\,, \\
G_M^\rho(Q^2=0) & = & \mu_\rho\,,\;
G_Q^\rho(Q^2=0) = Q_\rho\,.
\end{eqnarray}
It is readily seen that Eq.\,(\ref{chargenorm}) is a symmetry constraint.  One has $G_E(Q^2=0)=F_1(Q^2=0)$ and
\begin{equation}
\Lambda(K,Q) \stackrel{Q^2\to 0}{=} 2 K_\lambda\, {\cal P}^T_{\mu\alpha}(K) \, {\cal P}^T_{\alpha\nu}(K)\, F_1(0)\,.
\end{equation}
Using Eqs.\,(\ref{VWTI}), (\ref{PTQ2}), (\ref{LambdaGIA}), this becomes
\begin{eqnarray}
\nonumber
\lefteqn{ K_\lambda\, {\cal P}^T_{\mu\nu}(K) F_1(0) }\\
&=&
N_c E_{\rho}^2 {\rm tr}_{\rm D}\int\frac{d^4 q}{(2\pi^4)}
i\gamma_\nu\, \frac{\partial}{\partial K_\lambda} S(\ell+K) i\gamma_\mu\, S(\ell)\,.
\end{eqnarray}
The right-hand-side (rhs) is simply the analogue of Eq.\,(\ref{Ndef}) for the rainbow-ladder vector meson.  Hence, when $E_\rho$ is normalised according to Eq.\,(\ref{Erhonorm}) and so long as one employs a symmetry-preserving regularisation procedure, the rhs is equal to $K_\lambda\, {\cal P}^T_{\mu\nu}(K)$ and thus $F_1(0)=1$.

We compute the form factors using the formulae in Sec.\,\ref{rhoformulae}.  In Table~\ref{Table:radii} we report form factor radii, and the magnetic and quadrupole moments.  The comments following Eq.\,(\ref{rpidefn}) are also relevant to the magnitudes of the $\rho$-meson radii.
An interpretation of the ratio $r_\pi/r_\rho=0.80$ determined from the Table is complicated by the fact that we have consistently used the rainbow-ladder truncation; but in this case alone $F_\rho(P)=0$, whereas $F_\pi(P)\neq 0$ always and $F_\rho(P)\neq 0$ in all other truncations.  We observe therefore that $r_\pi = 0.51\,$fm if one artificially sets $F_\pi(P)=0$, in which case $r_\pi/r_\rho=0.92$.  Moreover, the DSE computation in Ref.\,\cite{Bhagwat:2006pu}, which employs a QCD-based interaction, produces $r_\pi/r_\rho=0.90$; and in combination, the more phenomenological DSE studies of Refs.\,\cite{Burden:1995ve,Hawes:1998bz} yield $r_\pi/r_\rho=0.92$.

\begin{figure}[t] 
\centerline{\includegraphics[clip,width=0.45\textwidth]{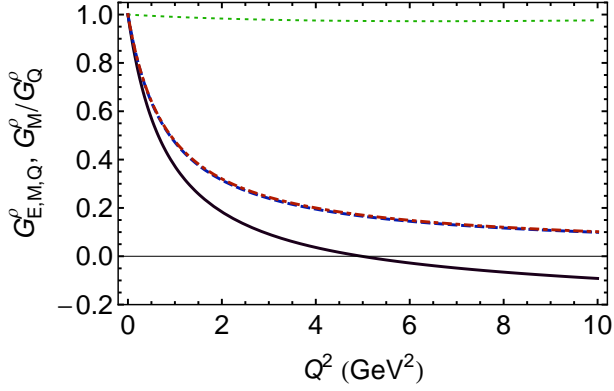}}
\caption{\label{FGAllrho} (Color online) \emph{Solid curve} -- $\rho$-meson electric form factor, $G_E^\rho(Q^2)$, which exhibits a zero at $Q^2=5.0\,$GeV$^2$.  (It is notable that $1-\frac{2}{3}\eta = 0 $ for $Q^2=6 m_\rho^2= 5.2\,$GeV$^2$.)  The \emph{dashed curve}, $G_M^\rho(Q^2)/\mu_\rho$, and \emph{dot-dashed curve}, $G_Q^\rho(Q^2)/{\cal Q}_\rho$, are almost indistinguishable, as emphasised by the \emph{dotted curve}, $[G_M^\rho(Q^2)/\mu_\rho]/[G_Q^\rho(Q^2)/{\cal Q}_\rho]$.
The charge radii, and magnetic and quadrupole moments are given in Table~\protect\ref{Table:radii}.
NB.\ All form factors exhibit a pole at $Q^2=-m_\rho^2$ because the quark-photon vertex is dressed as described in Sec.\,\protect\ref{sec:WTI}.
}
\end{figure}

Our computed $\rho$-meson electric form factor is plotted in Fig.\,\ref{FGAllrho}.  It displays a zero at $Q^2=5.0\,$GeV$^2$ and remains negative thereafter.  Given that the deuteron is a weakly-bound $J=1$ system, constituted from two fermions, and its electric form factor possesses a zero \cite{Kohl:2008zz}, it is unsurprising that $G_E^\rho(Q^2)$ exhibits a zero.  It is notable in addition that the deuteron's zero is located at $z_Q^{\rm D}:=\surd Q^2 = 0.8\,$GeV, so that
\begin{equation}
z_Q^{\rm D}r_{\rm D} \approx z_Q^\rho r_\rho^E\,,
\end{equation}
where $r_{\rm D}$ is the deuteron's radius.  An interpolation valid on $Q^2\in[-m_\rho^2,10\,{\rm GeV}^2]$ is
\begin{equation}
G_E^\rho(Q^2) \stackrel{\rm interpolation}{=} \frac{1-0.20\,Q^2}{1+1.15 \, Q^2-0.013\,Q^4}\,.
\end{equation}

In Fig.\,\ref{FGAllrho} we also depict the magnetic and quadrupole form factors of the $\rho$-meson,  both normalised by their values $Q^2=0$.  Notably, neither of these two form factors change sign: for $Q^2>-m_\rho^2$, $G_M^\rho(Q^2)$ is positive definite and $G_E^\rho(Q^2)$ is negative definite.  Furthermore, over this entire domain of $Q^2$, these form factors exhibit a very similar $Q^2$-dependence, which is made especially apparent via the dotted-curve in Fig.\,\ref{FGAllrho}.
Interpolations valid on $Q^2\in[-m_\rho^2,10\,{\rm GeV}^2]$ are
%
\begin{eqnarray}
G_M^\rho(Q^2) &\stackrel{\rm interpolation}{=} &
\frac{2.11+0.021\,Q^2}{1+1.15 \, Q^2-0.015\,Q^4}\,,\\
G_Q^\rho(Q^2) &\stackrel{\rm interpolation}{=} &
-\frac{0.85+0.038\,Q^2}{1+1.17 \, Q^2+0.014\,Q^4}\,.
\end{eqnarray}

The similar momentum-dependence of $G_M^\rho$ and $G_Q^\rho$ recalls a prediction in Ref.\,\cite{Brodsky:1992px}; namely,
\begin{equation}
\label{hillerratios}
G_E(Q^2):G_M(Q^2): G_Q(Q^2) \stackrel{Q^2\to \infty}{=} 1- \frac{2}{3}\eta  : 2 : -1
\end{equation}
in theories with a vector-vector interaction mediated via bosons propagating as $1/k^2$  at large-$k^2$.  Our computed ratio $r_{M/Q}:=G_M^\rho(Q^2)/G_Q^\rho(Q^2)$ conforms approximately with this prediction on a large domain of $Q^2$; e.g.,
\begin{equation}
\begin{array}{lcccc}
Q^2 & 0 & 10 & 10^2 & 10^3 \\
r_{M/Q}& -2.48 & -2.54 & -2.38 & -2.17
\end{array}\,.
\end{equation}
However, at $Q^2=10^4\,{\rm GeV}^2$, $r_{M/Q}=-1.28$.  Moreover, the remaining two ratios are always in conflict with the prediction; and closer inspection reveals that even the apparent agreement for $G_M^\rho(Q^2)/G_Q^\rho(Q^2)$ is accidental, since Eqs.\,(\ref{hillerratios}) are true if, and only if,
\begin{equation}
\label{hillerratiosF}
F_1(Q^2):F_2(Q^2): Q^2 F_3(Q^2) \stackrel{Q^2\to \infty}{=} 1  : -2 : 0\,;
\end{equation}
and none of these predictions are satisfied in our computation.

The mismatch originates, of course, with Eq.\,(\ref{njlgluon}) and the concomitant need for a regularisation procedure in which the ultraviolet cutoff plays a dynamical role.  If one carefully removes $\Lambda_{\rm uv} \to \infty$, Eqs.\,(\ref{hillerratiosF}) are recovered but at the cost of a logarithmic divergence in the individual form factors.
We conclude therefore that a vector-vector contact interaction cannot reasonably be regularised in a manner consistent with Eq.\,(\ref{hillerratios}).

In closing this subsection we reiterate that it is only in the rainbow-ladder truncation that $F_\rho(P)\equiv 0$.  Therefore in connection with the $\rho$-meson's form factors, material changes should be anticipated when proceeding beyond this leading-order truncation.

\begin{figure}[t] 
\centerline{\includegraphics[clip,width=0.45\textwidth]{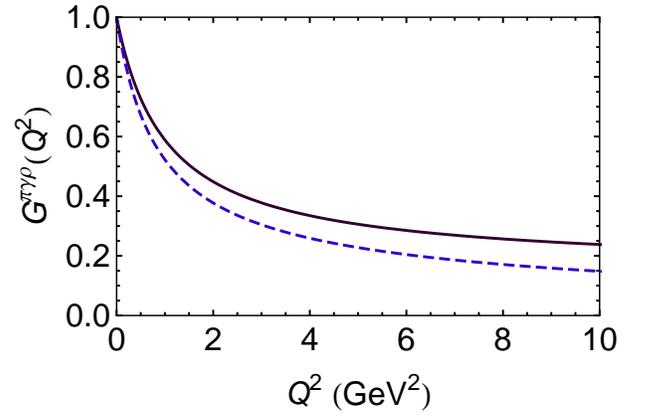}}
\caption{\label{FGpgr} (Color online) \emph{Solid curve} -- the full result for $G^{\pi\gamma\rho}(Q^2)$; and \emph{dashed curve} -- $G^{\pi\gamma\rho}(Q^2)$ obtained with $F_\pi(P)\equiv 0$.  Experimentally \protect\cite{Nakamura:2010zzi}, the partial width for $\rho^+ \to \pi^+ \gamma$ is $68\pm 7\,$keV, which corresponds to \protect\cite{Maris:2002mz} $g_{\pi\gamma\rho} = (0.74\pm0.05)\,m_\rho $.  This is in fair agreement with our computed result; viz., $g_{\pi\gamma\rho} = 0.63\,m_\rho$.}
\end{figure}

\subsection{\mbox{\boldmath $\rho$}-\mbox{\boldmath $\pi$} transition form factor}
This transition is closely related to the $\gamma^\ast \pi \gamma$ transition form factor, whose behaviour in connection with Eq.\,(\ref{njlgluon}) was analysed in Ref.\,\cite{Roberts:2010rn}.  The interaction vertex is expressed in Eq.\,(\ref{pirhotransition}) and defines a single form factor; viz.,
\begin{equation}
T_{\mu\nu}^{\pi\gamma\rho}(k_1,k_2) = \frac{g_{\pi\gamma\rho}}{m_\rho} \, \epsilon_{\mu\nu\alpha\beta}k_{1\alpha} k_{2\beta}\, G^{\pi\gamma\rho}(Q^2)\,,
\end{equation}
where $k_1^2=Q^2$, $k_2^2=-m_\rho^2$.  The coupling constant, $g_{\pi\gamma\rho}$, is defined such that $G^{\pi\gamma\rho}(Q^2=0)=1$; and explicit formulae for computing this form factor are provided in App.\,\ref{App:pgr}.

Our computed form factor is depicted in Fig.\,\ref{FGpgr}.  Naturally, because the quark-photon vertex is dressed (see Fig.\,\ref{qgammavertex}), the transition form factor exhibits a pole at $Q^2=-m_\rho^2$, which we have not displayed.  An interpolation valid on $Q^2\in[-m_\rho^2,10\,{\rm GeV}^2]$ is
\begin{equation}
G^{\pi\gamma\rho}(Q^2) \stackrel{\rm interpolation}{=}
\frac{1+0.37\,Q^2+0.024 Q^4}{1+1.29 \, Q^2+0.015\,Q^4}\,.\\
\end{equation}

In the neighbourhood of $Q^2=0$, the form factor is characterised by a radius-like length-scale; viz.,
\begin{equation}
\label{rpgr}
r_{\pi\gamma\rho}^2 := -6 \left. \frac{d}{dQ^2} G^{\pi\gamma\rho}(Q^2)\right|_{Q^2=0} = (0.46\,{\rm fm})^2,
\end{equation}
which is almost indistinguishable from both $r_\pi=0.45\,$fm in Table~\ref{Table:radii} and the anomaly interaction radius defined in Ref.\,\cite{Roberts:2010rn}; viz., $r_{\pi^0}^\ast=0.48\,$fm.  On the other hand
\begin{equation}
\label{rpgUV}
\lim_{Q^2\to \infty} G^{\pi\gamma\rho}(Q^2)= 0.11 \,, 
\end{equation}
owing to the presence of the pion's pseudovector component, a result in keeping with the pointlike nature of bound-states generated by a contact-interaction  \cite{GutierrezGuerrero:2010md,Roberts:2010rn}.

\begin{figure}[t] 
\centerline{\includegraphics[clip,width=0.45\textwidth]{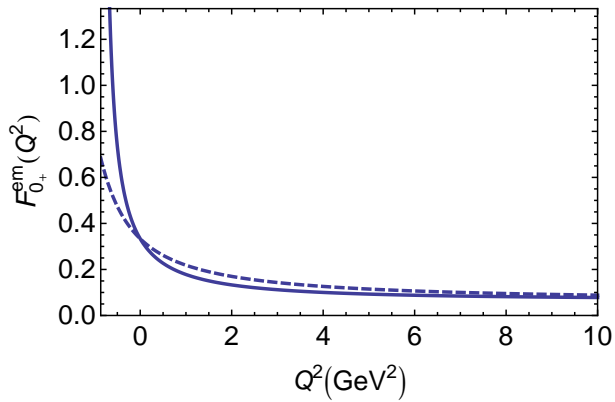}}
\caption{\label{Fem0} (Color online) \emph{Solid curve} -- full result for scalar-diquark elastic electromagnetic form factor; and \emph{dashed curve} -- result obtained without dressing the quark-photon vertex.  The computed mass of the diquark is $m_{qq_{0^+}}=0.776\,$GeV and the charge radius is given in Table~\protect\ref{Table:radii}.}
\end{figure}

\section{\mbox{\boldmath $0^+$}- and \mbox{\boldmath $1^+$}-diquark elastic and transition form factors}
\label{sec:formfactorsdiquarks}
\subsection{Scalar-diquark elastic form factor}
%
In the context of the interaction in Eq.\,(\ref{njlgluon}), a detailed discussion of the relationship between pseudoscalar- and vector-mesons and scalar- and axial-vector-diquark correlations may be found in Ref.\,\cite{Roberts:2011}.  Using the information provided therein, it is straightforward to show that in rainbow-ladder truncation the electromagnetic form factor of a scalar diquark is readily obtained from the expression for $F^{\rm em}_\pi(Q^2)$.  Namely,
\begin{equation}
F^{\rm em}_{0^+}(Q^2) = \frac{1}{3} \left. F^{\rm em}_{\pi}(Q^2) \right|^{(E_\pi,F_\pi)\to \sqrt{\frac{2}{3}}(E_{qq_{0^+}},F_{qq_{0^+}})}_{m_\pi\to m_{qq_{0^+}}},
\end{equation}
where the scalar-diquark Bethe-Salpeter amplitude is expressed via ($C=\gamma_2\gamma_4$ is the charge-conjugation matrix)
\begin{equation}
\Gamma_{qq_{0^+}}(P) C^\dagger = \gamma_5 \left[ i E_{qq_{0^+}}(P) + \frac{1}{M} \gamma\cdot P F_{qq_{0^+}}(P) \right].
\end{equation}

Our result for the scalar diquark elastic electromagnetic form factor is presented in Fig.\,\ref{Fem0}.  An interpolation valid on $Q^2\in [-m_\rho^2,10\,{\rm GeV}^2]$ is
\begin{equation}
F^{\rm em}_{0^+}(Q^2) \stackrel{\rm interpolation}{=} \frac{1}{3}
\frac{1+0.25\,Q^2+0.027\,Q^4}{1+1.27 Q^2 + 0.13 \,Q^4}\,.
\end{equation}
The normalisation is different but the momentum-dependence is similar to that of $F_\pi^{\rm em}$.  This is indicated, too, by the ratio of charge radii; viz., $r_{0^+}/r_\pi=1.08$, which may be compared to the value of $1.09$ obtained in Ref.\,\cite{Maris:2004bp} and contrasted with the value of $0.8$ in \cite{Bloch:1999ke}.  In the absence of the scalar-diquark Bethe-Salpeter amplitude's pseudovector component, $F_{qq_{0^+}}\equiv 0$, we find $r_{0^+}=0.51\,$fm; i.e., an increase of 6\%.

\begin{figure}[t] 
\centerline{\includegraphics[clip,width=0.45\textwidth]{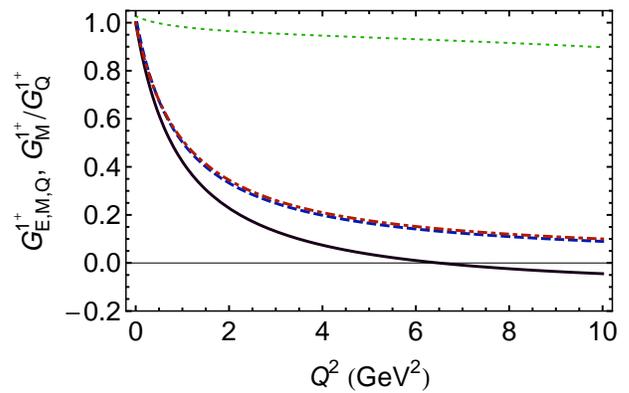}}
\caption{\label{Fem1} (Color online)
\emph{Solid curve} -- Pseudovector-diquark electric form factor, $G_E^{1^+}(Q^2)$, which exhibits a zero at $Q^2=6.5\,$GeV$^2$.  (In this case $1-\frac{2}{3}\eta = 0 $ for $Q^2=6 m_{1^+}^2= 6.7\,$GeV$^2$, given the computed mass of $1.06\,$GeV.)  The \emph{dashed curve}, $G_M^{1^+}(Q^2)/\mu_{1^+}$, and \emph{dot-dashed curve}, $G_Q^{1^+}(Q^2)/{\cal Q}_{1^+}$, are almost indistinguishable, as emphasised by the \emph{dotted curve}, $[G_M^{1^+}(Q^2)/\mu_\rho]/[G_Q^{1^+}(Q^2)/{\cal Q}_{1^+}]$.
The charge radii, and magnetic and quadrupole moments are given in Table~\protect\ref{Table:radii}.
NB.\ All form factors exhibit a pole at $Q^2=-m_\rho^2$ because the quark-photon vertex is dressed as described in Sec.\,\protect\ref{sec:WTI}.
}
\end{figure}

\subsection{Pseudovector-diquark elastic form factors}
From the above observations it will be apparent that the rainbow-ladder results for the $\{u d\}$ axial-vector diquark elastic form factors may be obtained directly from those of the $\rho$-meson through the substitutions
\begin{equation}
F^{\rm em}_{1^+_{\{ud\}},j}(Q^2) = \frac{1}{3} \left. F_{j}(Q^2) \right|^{E_\rho\to \sqrt{\frac{2}{3}} E_{qq_{1^+}}}_{m_\pi\to m_{qq_{1^+}}}\,.
\end{equation}
The momentum-dependence of the form factors for the $\{uu\}$ and $\{dd\}$ correlations is identical but in these cases the normalisations are, respectively, $\frac{4}{3}$ and $-\frac{2}{3}$.

We depict the axial-vector diquark form factors in Fig.\,\ref{Fem1}.  They are similar to but distinguishable from those of the $\rho$-meson, falling-off a little less rapidly owing to the larger mass of the axial-vector diquark.  Interpolations valid on $Q^2\in[-m_\rho^2,10\,{\rm GeV}^2]$ are
%
%
\begin{eqnarray}
G_E^{1^+}(Q^2) &\stackrel{\rm interpolation}{=} &
\frac{1-0.16\,Q^2}{1+1.17 \, Q^2 + 0.012 \,Q^4}\,,\\
G_M^{1^+}(Q^2) &\stackrel{\rm interpolation}{=}&
\frac{2.13-0.19\,Q^2}{1+1.07 \, Q^2 - 0.10 \,Q^4}\,,\\
G_Q^{1^+}(Q^2) &\stackrel{\rm interpolation}{=}&
-\frac{0.81-0.029\,Q^2}{1+1.11 \, Q^2 - 0.054 \,Q^4}\,,
\end{eqnarray}
from which the particular pseudovector diquark form factors are obtained after multiplication by the appropriate charge factors, listed in Table~\ref{Table:radii}.


\subsection{\mbox{\boldmath $1^+$}-\mbox{\boldmath $0^+$} diquark transition form factor}
Owing to the flavour structure of the scalar diquark, this transition can only involve the $\{ud\}$ axial-vector diquark.  It is described by a single form factor, which can be introduced through
\begin{equation}
T_{\mu\nu}^{0^+ \gamma 1^+}(k_1,k_2) = \frac{1}{3} \frac{g_{0^+ \gamma 1^+}}{m_{qq_{1^+}}} \, \epsilon_{\mu\nu\alpha\beta}k_{1\alpha} k_{2\beta}\, G^{0^+ \gamma 1^+}(Q^2)\,,
\end{equation}
and one may readily determine that in rainbow-ladder truncation
\begin{eqnarray}
\nonumber
\lefteqn{G^{0^+ \gamma 1^+}(Q^2)}\\
& =&
\left. G^{\pi\gamma\rho}(Q^2) \right|^{(E_\pi,F_\pi,E_\rho)\to \sqrt{\frac{2}{3}}(E_{qq_{0^+}},F_{qq_{0^+}},E_{qq_{1^+}})}_{m_\pi\to m_{qq_{0^+}},m_\rho\to m_{qq_{1^+}}}\!\!\!\!.
\end{eqnarray}

\begin{figure}[t] 
\centerline{\includegraphics[clip,width=0.45\textwidth]{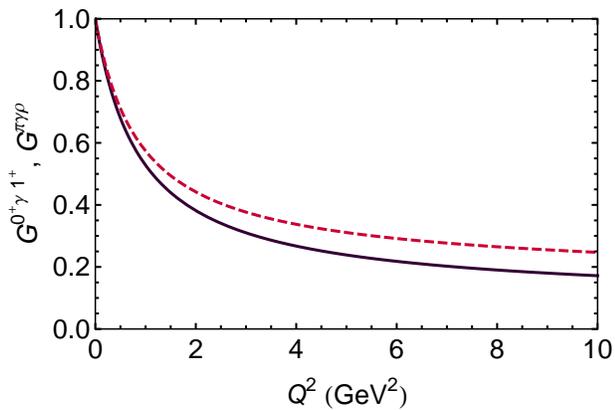}}
\caption{\label{F10g} (Color online)
\emph{Solid curve} -- momentum-dependence of full result for axial-vector--scalar-diquark transition form factor, $G^{0^+\! \gamma 1^+}(Q^2)$; and \emph{dashed curve} -- result for $G^{\pi\gamma\rho}(Q^2)$ in Fig.\,\protect\ref{FGpgr}. The different rates of evolution are typical of meson cf.\ diquark form factors computed herein.
Note that $e_{\{ud\}}g_{0^+\gamma 1^+} m_{qq_{1^+}} = e_{\{ud\}} 0.74 = 0.25$.
}
\end{figure}

Computation of the form factor is straightforward and the result is depicted in Fig.\,\ref{F10g}.  An interpolation valid on $Q^2 \in [-m_\rho^2,10\,{\rm GeV}^2]$ is
\begin{equation}
G^{0^+ \gamma 1^+}(Q^2) \stackrel{\rm interpolation}{=}
\frac{1+0.10\,Q^2}{1+1.073\,Q^2}\,.
\end{equation}
The associated transition radius is
\begin{equation}
r_{0^+\gamma 1^+} = 0.48\,{\rm fm},
\end{equation}
which is 5\% larger than $r_{\pi\gamma \rho}$ in Eq.\,(\ref{rpgr}), and
\begin{equation}
\lim_{Q^2\to \infty} G^{0^+\gamma 1^+}(Q^2)= 0.049 \,, 
\end{equation}
just under one-half of the value in Eq.\,(\ref{rpgUV}).

\section{Epilogue}
\label{sec:epilogue}
We described a unified Dyson-Schwinger equation treatment of static and electromagnetic properties of pseudoscalar and vector mesons, and scalar and axial-vector diquark correlations based upon a vector-vector contact-interaction.  Isospin symmetry was assumed, with $m_u=m_d=m=7\,$MeV producing a physical pion mass; and two parameters were used to define the gap- and Bethe-Salpeter.  In a comparison with relevant static quantities, we recorded a value of 13\% for the overall root-mean-square relative-error.

A basic motivation for our study is the need to document a comparison between the electromagnetic form factors of mesons and those diquarks which play a material role in nucleon structure because this is an important step toward a unified description of meson and baryon form factors based on a single interaction.  A notable feature of our results, therefore, is the large degree of similarity between related form factors.  For example, we find that it would be a good practical approximation to assume equality of related radii: $r_{0^+}\approx r_\pi$ and $r_{1^+} \approx r_\rho$.

As has previously been observed, a fully-consistent treatment of the contact interaction produces a pion electromagnetic form factor that approaches a nonzero constant value at large spacelike momenta.  On the other hand, owing to a peculiarity of the rainbow-ladder truncation, which prevents the appearance at this order of a tensor component for the $\rho$-meson produced by a contact interaction, the $\rho$-meson form factors approach zero at large spacelike momenta.  This accident means that a comparison with QCD-based DSE calculations can meaningfully be interpreted.  In a comparison with the most sophisticated such study, the form factors produced by the contact interaction are harder although the ratio $r_\pi/r_\rho^E$ is similar.  Moreover, the contact interaction's simplicity allows one to readily compute the $\rho$-meson form factors at arbitrarily large spacelike $Q^2$ and expose a zero in the electric form factor at $z_Q^2 \approx 6 m_\rho^2$.  Notably, $r_{\rm D} z_Q^{\rm D}\approx r_\rho^E z_Q^\rho$, where $r_{\rm D}$ and $z_Q^{\rm D}$ are, respectively, the deuteron's radius and the location of the zero in its electric form factor.  The $\rho$-meson's magnetic and quadrupole form factors are positive- and negative-definite, respectively.  We reiterate that the behaviour of all pseudovector-diquark form factors is semiquantitatively the same.

At the core of our analysis is a symmetry-preserving treatment of a vector-vector contact interaction.  This has now been used in the completely-consistent computation of the hadron spectrum, and meson and diquark form factors.  The foundation has thus been laid for the computation of baryon elastic and transition form factors, which will provide information that is crucial for the use of experimental data on such observables as a tool for charting the nature of the quark-quark interaction at long-range \cite{Aznauryan:2009da}.

\begin{acknowledgments}
We acknowledge valuable discussions with L.~Chang, I.\,C.~Clo\"et, C.~Hanhart and S.\,M.~Schmidt.
This work was supported by:
the U.\,S.\ Department of Energy, Office of Nuclear Physics, contract no.~DE-AC02-06CH11357;
Forschungszentrum J\"ulich GmbH;
the Department of Energy's Science Undergraduate Laboratory Internship programme;
CIC and CONACyT grants, under project nos.\ 4.10 and 46614-I;
and the Project of Knowledge Innovation Program of the Chinese Academy of Sciences, Grant No.\ KJCX2.YW.W10.
\end{acknowledgments}

\appendix
\section{Form Factor Formulae}
This appendix is a repository for the formulae we have used to compute the form factors.  
\subsection{Elastic pion form factor}
\label{app:FpiQ2}
\begin{eqnarray}
\nonumber
F_\pi^{\rm em}(Q^2) & = & P_T(Q^2) \left[ E_\pi^2 T_{\pi,EE}^{\rm em}(Q^2) \right. \\
&+ & \left. E_\pi F_\pi T_{\pi,EF}^{\rm em}(Q^2)+ F_\pi^2 T_{\pi,FF}^{\rm em}(Q^2) \right],
\end{eqnarray}
where $P_T(Q^2)$ is given in Eq.\,(\ref{PTQ2}) and
\begin{widetext}
\begin{eqnarray}
T_{\pi,EE}^{\rm em} & = &  \frac{3}{4\pi^2} \left[ \int_0^1 \! d\alpha\,  \overline{\cal C}^{\rm iu}_1(\omega(M^2,\alpha,Q^2))+
 2 m_\pi^2 \int_0^1 \! d\alpha\, d\beta\, \alpha \,
\overline{\cal C}_2^{\rm iu}(\omega_2(M^2,\alpha,\beta,Q^2,m_\pi^2))\right],\\
T_{\pi,EF}^{\rm em} & = & \frac{3}{2\pi^2} \left[ - \int_0^1 \! d\alpha\,  \overline{\cal C}^{\rm iu}_1(\omega(M^2,\alpha,Q^2))+  \int_0^1 \! d\alpha\, d\beta\, \alpha \,(\alpha Q^2 - 2 m_\pi^2)\,
\overline{\cal C}_2^{\rm iu}(\omega_2(M^2,\alpha,\beta,Q^2,m_\pi^2))\right],\\
T_{\pi,FF}^{\rm em} & = & -\frac{3}{2\pi^2}\frac{1}{M^2} \int_0^1\! d\alpha d\beta\,
\alpha \left[ {\cal A}(\alpha,Q^2,m_\pi^2) \,
\overline{\cal C}_1^{\rm iu}(\omega_2(M^2,\alpha,\beta,Q^2,m_\pi^2)) \right. \\
&+& \left.
[{\cal B}(M^2,\alpha,\beta,Q^2,m_\pi^2) -
{\cal A}(\alpha,Q^2,m_\pi^2)\,\omega_2(M^2,\alpha,\beta,Q^2,m_\pi^2) ] \,
\overline{\cal C}_2^{\rm iu}(\omega_2(M^2,\alpha,\beta,Q^2,m_\pi^2))
\right],
\end{eqnarray}
with
\begin{equation}
{\cal B}(M^2,\alpha,\beta,Q^2,m_\pi^2) =
\alpha Q^2 M^2 + M^2 m_\pi^2(\alpha-2)+ \alpha m_\pi^2 (\alpha Q^2 [1-\alpha-2 \beta(1-\beta)+3 \alpha\beta(1-\beta)]-(1-\alpha)^2 m_\pi^2),
\end{equation}
\end{widetext}
\begin{equation}
{\cal A}(\alpha,Q^2,m_\pi^2)=  -\frac{1}{2}\, \alpha\, Q^2 + \frac{1}{2} m_\pi^2 (2-3\alpha),
\end{equation}
\begin{eqnarray}
\nonumber \omega_2(M^2,\alpha,\beta,Q^2,m_\pi^2) &=&
M^2 + Q^2 \alpha^2 \beta (1 - \beta) \\
&& - \alpha (1 - \alpha) m_\pi^2 ,
\end{eqnarray}
where
${\cal C}^{\rm iu}(z)$ is defined after Eq.\,(\ref{gapactual}); ${\cal C}_1^{\rm iu}(z)$ and $\omega(M^2,\alpha,Q^2)$ in Eqs.(\ref{eq:omega}), (\ref{eq:C1}); $\overline{\cal C}_1^{\rm iu}(z)$ after Eq.\,(\ref{PTQ2}); and
\begin{equation}
{\cal C}_2^{\rm iu}(z) = z^2 {\cal C}^{\prime\prime}(z)= \frac{z}{2} \left({\rm e}^{-z r_{\rm uv}^2} - {\rm e}^{-z r_{\rm ir}^2}\right)
\end{equation}
with $\overline{\cal C}_2^{\rm iu}={\cal C}_2^{\rm iu}(z)/z^2$.

\subsection{Elastic \mbox{\boldmath $\rho$}-meson form factors}
\label{rhoformulae}
In generalised impulse approximation the $\rho$-$\gamma$ vertex in Eq.\,(\ref{Lambdarho}) becomes
\begin{eqnarray}
\nonumber
\Lambda_{\lambda,\mu\nu} & = & 2 N_c {\rm tr}_{\rm D}\int\frac{d^4 q}{(2\pi^4)}
E_{\rho}(-p^f)\gamma_\nu^{\rm T}\, S(q + p^f) \, P_T(Q^2) i\gamma_\lambda\, \\
&& \times S(q + p^i) E_{\rho}(p^i) \gamma_\mu^{\rm T}\, S(q)\,, \label{LambdaGIA}
\end{eqnarray}
where $E_\rho$ is the canonically-normalised $\rho$-meson Bethe-Salpeter amplitude.  Explicit expressions for the scalar functions $F_{1,2,3}(Q^2)$ can be obtained via contraction with any three sensibly chosen projection operators; and the subsequent use of Feynman parametrisations yields
\begin{eqnarray}
\nonumber
F_i(Q^2) & = & \frac{3}{4\pi^2}E_\rho^2 \int_0^1\! d\alpha d\beta\,
\alpha \left[ {\cal A}_i \,
\overline{\cal C}_1^{\rm iu}(\omega_2) \right. \\
&&+ \left.
[{\cal B}_i - {\cal A}_i\,\omega_2 ] \,
\overline{\cal C}_2^{\rm iu}(\omega_2)
\right],
\end{eqnarray}
%
where ${\cal F}_i={\cal F}_i(M^2,\alpha,\beta,Q^2,m_\rho^2)$, ${\cal F}_i={\cal A}_i$, ${\cal B}_i$; viz.,
\begin{eqnarray}
{\cal A}_1 & = & 2-\alpha \,,\\
{\cal A}_2 & = & \frac{m_\rho^2 (\alpha (10 \beta - 7)-4)+ Q^2 \alpha (2 \beta-1)}{2 m_\rho^2}\,,
\end{eqnarray}
\begin{eqnarray}
{\cal A}_3 & = & \frac{2 \alpha (1-2\beta) (5 m_\rho^2 + Q^2)}{4 m_\rho^2+Q^2}\,,\\
\nonumber
{\cal B}_1 & = & 2 \left[ M^2 (2-\alpha) + m_\rho^2 \alpha (1-\alpha)^2 \right.\\
& & \left. - \alpha^2 \beta (2-\alpha) (1-\beta) Q^2\right]\,,
\end{eqnarray}
\begin{widetext}
\begin{eqnarray}
\nonumber
m_\rho^2 \, {\cal B}_2 & = &
 m_\rho^2
[ M^2 (-4 - 7 \alpha + 10 \alpha \beta) -
m_\rho^2(-1+\alpha) \alpha (1-7\alpha - 6 \beta + 10 \alpha \beta )]  \\
&&  + \alpha [M^2 (-1+2\beta) + m_\rho^2 \alpha (-1+2 \beta + \alpha [1+\beta - 5 \beta^2+ 2 \beta^3]) Q^2]\,,\\
\nonumber
(4 m_\rho^2 +Q^2) {\cal B}_3 & = &
4 \alpha
\left[m_\rho^2
(5 M^2 (1-2\beta) + m_\rho^2 (-1+\alpha) [3 - 6 \beta + \alpha (-5-6\beta+16 \beta^2)])
\right. \\
&& \left. + (M^2 (1-2\beta) - m_\rho^2 \alpha [-1+\alpha-2\beta + 3\alpha\beta + 4\beta^2-7\alpha\beta^2+2 \alpha\beta^3]) Q^2\right]\,.
\end{eqnarray}
\end{widetext}

\subsection{Vector-pseudoscalar transition form factor}
\label{App:pgr}
The interaction vertex describing the $\pi$-$\rho$ transition
\begin{eqnarray}
T_{\mu\nu}^{\pi\gamma\rho}(k_1,k_2) &=& \frac{g_{\pi\gamma\rho}}{m_\rho} \, \epsilon_{\mu\nu\alpha\beta}k_{1\alpha} k_{2\beta}\, G^{\pi\gamma\rho}(Q^2)\\
\nonumber
& = & {\rm tr}_{\rm D}\!\!\! \int\!\!\!\frac{d^4 \ell}{(2\pi)^4} \, \Gamma_\pi(-P) S(\ell_2) \,P_T(Q^2)\,i\gamma_\mu \\
&& \times \, S(\ell_{12}) \, i  \Gamma^\rho_\nu(k_2)\,S(\ell_1) \,,
 \label{pirhotransition}
\end{eqnarray}
where the incoming $\rho$-meson has momentum $k_2$, the photon has momentum $k_1=Q$, the outgoing pion has momentum $P=(k_1+k_2)$; and $\ell_{1}=\ell-k_1$, $\ell_{2} = \ell + k_2$, $\ell_{12}=\ell - k_1 + k_2$.  In this instance the kinematic constraints are
\begin{equation}
k_1^2=-m_\rho^2\,,\; k_2^2=Q^2\,, \; 2\, k_1\cdot k_2= m_\rho^2- m_\pi^2- Q^2\,.
\end{equation}

Given the structure of the pion's Bethe-Salpeter amplitude, one may write
\begin{equation}
G^{\pi\gamma\rho}(Q^2) = G_E^{\pi\gamma\rho}(Q^2) + G_F^{\pi\gamma\rho}(Q^2)  \,,
\end{equation}
wherein
\begin{eqnarray}
\label{GEprg}
G_E^{\pi\gamma\rho}(Q^2) &=& \frac{E_\pi {\cal E}_\rho }{2\pi^2}M
\int_0^1 d\alpha d\beta \, \alpha \, \bar {\cal C}_2^{\rm ir}(\omega_3)\,,\\
\nonumber \hat G_F^{\pi\gamma\rho}(Q^2) &=& -\frac{F_\pi {\cal E}_\rho }{4\pi^2}\frac{1}{M}
\int_0^1 d\alpha d\beta \, \alpha\,
\left[ f_1^{\pi\gamma\rho} \bar {\cal C}_1^{\rm ir}(\omega_3) \right.\\
&&
\left. + (f_0^{\pi\gamma\rho} - \omega_3 \,f_1^{\pi\gamma\rho} ) \bar {\cal C}_2^{\rm ir}(\omega_3)
\right],
\label{GFprg}
\end{eqnarray}
with
\begin{eqnarray}
\nonumber
\omega_3 &:= & \omega_3(M^2,\alpha,\beta,m_\rho^2,m_\pi^2,Q^2)\\
\nonumber
&=& M^2 - \alpha \left[ \alpha \beta (1 - \beta) m_\pi^2
 + (1-\alpha) (1-\beta) m_\rho^2 \right.\\
&& \left. - (1-\alpha)\beta Q^2 \right]
\end{eqnarray}
and
\begin{eqnarray}
f_1^{\pi\gamma\rho} & = & 2- 3 \alpha\,,\\
\nonumber
f_0^{\pi\gamma\rho} & = & (2-\alpha) (M^2+\alpha^2 \beta (1-\beta) m_\pi^2) \\
&&  +(1-\alpha)^2 ( \alpha (1-\beta) m_\rho^2 - \alpha \beta Q^2)\,.\\
&& \nonumber
\end{eqnarray}

The vertex in Eq.\,(\ref{pirhotransition}) is intimately connected with the Abelian anomaly, which describes the process $\pi^0 \to \gamma\gamma$ and associated transition form factors.  The manner by which all aspects of the anomaly may be reproduced in the model we're considering is detailed in Secs.\,III.A and III.B.2 of Ref.\,\cite{Roberts:2010rn}.  In the present context, consistency with the anomaly
requires that in Eqs.\,(\ref{GEprg}), (\ref{GFprg}), ${\cal E}_\rho = E_\rho/{\cal N}_{\pi\gamma\gamma}$, with ${\cal N}_{\pi\gamma\gamma}$ defined such that $G^{\pi\gamma\gamma}(Q^2=0)= 1/2$, and
$G_F^{\pi\gamma\rho}(Q^2)=\hat G_F^{\pi\gamma\rho}(Q^2)-\hat G_F^{\pi\gamma\rho}(0)$.
Both modifications are necessary in order to correct for the dynamical role played by the ultraviolet cutoff in a contact-interaction theory.

\end{document}